\documentstyle[eqsecnum,aps,psfig]{revtex}

\def\etal{{\it et\thinspace al.}\ }
\def\eion{{(e~+~ion)}\ }
\def\efe18{{(e~+~Fe~XVIII)}\ }
\def\efe1817{{(e~+~Fe~XVIII) $\rightarrow$ Fe~XVII}\ }
\def\be{\begin{equation}}
\def\ee{\end{equation}}
\begin{document}
\title{Relativistic close coupling calculations for photoionization
and recombination of Ne-like Fe XVII}
\author{Hong Lin Zhang}
\address{
Applied Theoretical \& Computational Physics Division,\\
Los Alamos National Laboratory, Los Alamos, NM 87545\\
}
\author{Sultana N. Nahar and Anil K. Pradhan}
\address{
Department of Astronomy, The Ohio State University, Columbus, Ohio
43210\\
}
\date{\today}
\maketitle
\begin{abstract}
Relativistic and channel coupling effects in photoionization and
unified electronic recombination of Fe~XVII are demonstrated with
an extensive 60-level close coupling calculation using the Breit-Pauli
R-matrix method. A multi-configuration eigenfunction expansion up to the
$n$ = 3 levels of the
core ion Fe~XVIII is employed with 5 spectroscopic configurations
$2s^2p^5, 2s2p^6, 2s^22p^4 \ 3s, 3p, 3d$, and a number of correlation 
configurations. The unified \eion recombination calculations
for \efe1817 include both the non-resonant
and resonant recombination (`radiative' and `dielectronic
recombination' -- RR and DR). Photoionization  and \eion recombination
calculations are carried out for the total and the 
level-specific cross
sections, including the ground and several hundred excited
bound levels of Fe~XVII (up to fine structure levels with $n$ = 10). The
low-energy and the high energy cross sections are compared from: (i) a
3-level calculation including only the $2s^2p^5 \ \
(^2P^o_{1/2,3/2})$ and $2s2p^6 \ \ (^2S_{1/2})$ levels of Fe~XVIII, and
(ii) the first 60-level calculation with $\Delta n > 0$ coupled
channels. Strong channel coupling effects are  demonstrated
throughout the energy ranges considered, in
particular via giant photoexcitation-of-core (PEC) resonances
due to L-M shell dipole transition arrays $2p^5 \rightarrow 2p^4 \ 3s, 3d$ in
Fe~XIII that enhance effective cross sections by orders of
magnitude. Comparison is made with previous theoretical and 
experimental works on photoionization and recombination that considered
the relatively small low-energy region (i), and the weaker $\Delta n = 0$ 
couplings. While the simpler
3-level results describe the near-threshold photoionization and
recombination, they are inadequate for practical applications
that also require the higher energy cross sections for modeling ionization
balance of Fe~XVII in laboratory and astrophysical plasmas. The present
60-level results should provide reasonably complete and accurate
datasets for both photoionization and \eion recombination of Fe~XVII.

\end{abstract}

\section{INTRODUCTION}

 Laboratory, astrophysical, and theoretical studies of Fe~XVII
are of considerable interest as it is a prime constituent in high temperature
plasmas responsible for strong X-ray emission \cite{letal,bk99,metal}. 
A number of atomic
processes need to be considered in detail, primarily: electron impact 
excitation, photoionization, \eion recombination, and radiative transitions.
Large-scale atomic calculations are in progress for all of these
processes in Fe~XVII under the Iron Project and related works 
\cite{ip,symp,ipwww}, in
extended energy ranges suitable for practical applications. While
electron impact excitation is an independent part of this effort, in this work
we describe photoionization and \eion recombination of Fe~XVII.
 The coupled channel approximation including 
relativistic effects for many channel systems can be very involved
owing to many infinite series of resonance structures converging on to the
various excited levels of the core ion.
Whereas the relativistic and coupling effects have been studied
previously, all such
theoretical and experimental studies of Fe~XVII photoionization (e.g.
Haque \etal \cite{hetal} and Mohan \etal \cite{metal})
have been limited to the ground state and the relatively small energy 
range spanned by core excitations
within the $n$ = 2 complex of the residual ion Fe~XVIII 
comprising of 3 fine structure levels up to about 132 eV, i.e.

\be
 h\nu + {\rm Fe~XVII} \ (2s^22p^6 \ ^1S_0) \longrightarrow e + {\rm Fe~XVIII} 
\ (2s^22p^5 \ ^2P^o_{3/2,1/2}), \ 2s2p^6 \ (^2S_{1/2}).
\ee

 Although the near-threshold behavior of photoionization and
recombination cross sections (Pradhan \etal \cite{petal}) is physically 
interesting, it is
inadequate for practical applications that require the cross sections to
be calculated up to high energies typical of the variety of conditions
where L-shell ions are abundant. Purely photoionized plasmas, such as in
H~II regions, planetary nebulae, novae etc., are typically low
temperature, but coronal plasmas cover a much wider range \cite{tk95}. 
For example, the temperature of maximum
abundance of Fe~XVII in the coronal ionization equilibrium is about 4 $\times
10^6$ K \cite{ar92}. Furthermore, in astrophysical objects such as the warm 
absorber
ionized gas thought to surround the central black hole in active galactic
nuclei, the plasma is most likely of a composite nature since
most ionization states of several elements are observed (e.g. \cite{letal}).
High accuracy throughout the energy range of practical importance is
therefore essential.

 As the (e~+~ion) recombination  is unified in nature, it is
theoretically
desirable to consider the non-resonant and resonant processes (RR and
DR) together. A unified theoretical formulation has been developed
\cite{np94,zetal}
including relativistic
fine structure \cite{zp97},
and used to compute cross sections and rates for many
atomic systems, such as the K-shell systems C~IV -- C~V and Fe~XXIV --
Fe~XXV
of interest in X-ray spectroscopy \cite{n00a,n00b}.
The unified results may be directly compared with experimental results, without 
the
need to separate RR and DR. In this paper  we present details of the
low-energy results for Fe~XVII and show that not only are the unified cross
sections and rates in good agreement with experiment, but also
illustrate how the unified calculations avoid the basic inconsistency
and incompleteness of photoionization and recombination data for
modeling of laboratory and astrophysical plasma sources.

 The present report describes in detail the 3-level and the 60-level
close coupling calculations, with a discussion of the relativistic and
coupling effects and comparison with earlier 3-level theoretical and
experimental data for photoionization and recombination. While photoionization
and recombination are usually considered separately, we exemplify and
emphasize the underlying physical unity, via detailed balance, between the 
two processes as naturally treated in the close coupling method.

\section{THEORY}

 Photoionization and \eion recombination may both be considered using
{\it identical} coupled channel wavefunction expansion,

\begin{equation}
\Psi(E) = A \sum_{i} \chi_{i}\theta_{i} + \sum_{j} c_{j} \Phi_{j},
\end{equation}

where $\Psi$ represents a ($N$+1)-electron bound or continuum state depending 
on $E<$0 or $E>$0, expressed in terms of the $N$-electron residual core ion 
eigenfunctions.
The $\chi_{i}$ is the target wavefunction in a specific state
$S_iL_i\pi_i$ or $J_i\pi_i$, and $\theta_{i}$ is the wavefunction for the
($N$+1)-th electron in a channel labeled as
$S_iL_i(J_i)\pi_ik_{i}^{2}\ell_i(SL\pi \ {\rm or}\ J\pi))$;
$k_{i}^{2}$ being its incident kinetic energy. $\Phi_j$'s are the correlation
functions of the ($N$+1)-electron system that account for short range
correlation and the orthogonality between the continuum and the bound
orbitals. The R-matrix method \cite{burke,betal},
and its relativistic
Breit-Pauli extension \cite{st81}, enables a solution for
the total $\Psi$, with a suitable expansion over the $\chi_i$.
The Breit-Pauli R-matrix (BPRM) method has been
extensively employed for electron impact excitation under the
Iron Project \cite{ip,rm}.
The extension of the BPRM formulation to unified electronic
recombination \cite{zetal,n00b}, and theoretically self-consistent
calculations of photoionization and recombination is sketched below.

 Recombination of an incoming electron to the target ion may occur through
non-resonant, background continuum, usually referred to as
radiative recombination (RR),

\begin{equation}
e + X^{++} \rightarrow  h\nu + X^+,
\end{equation}

\noindent
which is the inverse process of direct photoionization, or through the
two-step recombination process via autoionizing resonances, i.e.
dielectronic recombination (DR):

\begin{equation}
e + X^{++} \rightarrow (X^+)^{**}  \rightarrow  \left\{
\begin{array}{c}  e + X^{++} \\   h\nu + X^+ \end{array} \right. ,
\end{equation}

\noindent
where the incident electron is in a
quasi-bound doubly-excited state which leads either to (i)
autoionization,
a radiation-less transition to a lower state  of the ion and the free
electron, or to (ii) radiative stabilization predominantly
via decay of the ion core,
usually to the ground state, and the bound electron.

In the unified treatment the photoionization cross sections,
$\sigma_{\rm PI}$,
of a large number of low-$n$ bound states -- all possible states with
$n \leq n_{\rm max}\sim 10$ -- are obtained in the close coupling (CC)
approximation
as in the Opacity Project \cite{symp}. Coupled channel calculations
for $\sigma_{\rm PI}$ include both
the background and the resonance structures (due to the doubly excited
autoionizing states) in the cross sections. 
The recombination cross section,
$\sigma_{\rm RC}$, is related to $\sigma_{\rm PI}$, through detailed
balance
(Milne relation) as

\begin{equation}
\sigma_{\rm RC}(\epsilon) =
{\alpha^2 \over 4} {g_i\over g_j}{(\epsilon + I)^2\over
\epsilon}\sigma_{\rm PI}
\end{equation}
in Rydberg units; $\alpha$ is the fine structure constant, $\epsilon$ is
the photoelectron energy, and $I$ is the ionization potential.

 Resonant and non-resonant electronic recombination takes place into
an infinite number of bound levels of the (e~+~ion) system. These are
divided into two groups: (A) the low-$n$ ($n \leq n_o \approx$ 10)
levels, considered via
detailed close coupling calculations for photorecombination, with
highly resolved delineation of autoionizing resonances, and (B) the
high-$n$ ($n_o \le n \leq \infty$) recombining levels via DR,
neglecting the background. In previous works (e.g. \cite{zetal})
it has been shown that in
the energy region corresponding to (B), below threshold for DR,
the non-resonant contribution is negligible.
The DR cross sections converge on to the
electron impact excitation cross section
at threshold ($n \rightarrow \infty$), as
required by unitarity, i.e. conservation of photon and electron fluxes.
This theoretical limit is an important check on the calculations, and
enables a determination of field ionization of rydberg levels of
resonances contributing to DR.

 The ab initio method outlined above is a theoretically and
computationally unified treatment based on the close coupling
approximation. Recombination involves an infinite number of recombined
bound states, and several
infinite series of resonances. In principle, the unified method may be
used for photoionization/photorecombination of arbitrarily high $n,l,J$
levels. However, in practice
approximations may be made for sufficiently high quantum numbers.
Background recombination is negligible, and DR
dominates, usually for $n_{max} \ge $ 10. Similarly, background (non-resonant)
cross
sections may be accurately obtained using hydrogenic approximation for
$n ,l$ levels with $n > 10$. But there is nothing particular about 
$n_{max}$ = 10,
and any larger or lower value may be used provided the approximations are
verifiably valid, as has been shown in our previous works and is done in
the present calculations. For example, $n$ = 18 -- 22+ resonances in the 
present work are fully delineated using group (A) photorecombination 
calculations, and not DR. Thus the use of these approximations
does not result in any significant error, or loss of generality, and
does not detract from the main part of the calculations that are a unified
representation of the non-resonant and resonant recombination (RR
and DR), including any interference effects between the two. The present DR
calculations use an extension of the precise theory by Bell and Seaton 
\cite{bs},
based on multi-channel quantum defect theory, that is very accurate
for high-$n$ (correspondence between
photorecombination and DR is established in our previous work \cite{zetal}).
Finally, all close
coupling scattering and photoionization calculations employ a
``top-up'' procedure for high partial waves, and approximations for
high-$n$ resonances below
Rydberg series limits as  $n \rightarrow \infty$ (e.g. ``Gailitis
averaging''). Such procedures are routinely implemented in large-scale
calculations in the Opacity Project (\cite{symp} and references therein) and
the Iron Project R-matrix calculations (\cite{ipwww}) 
that the unified method for \eion recombination is based upon.

\section{COMPUTATIONS}

 The complete wavefunction expansion entails the 60 fine structure
levels of Fe~XVIII given in Table 1. These are obtained from an optimized
configuration-interaction (CI) type calculation using the code SUPERSTRUCTURE
\cite{ss}. The configuration set is divided into the 5 {\it spectroscopic}
configurations $2s^2 2p^5, 2s2p^6, 2s^22p^4 \ 3s, 3p, 3d$
that dominate the 60 core level
wavefunctions, and {\it correlation} configurations, 
$2sp^5\ 3s, 3p, 3d$, $2p^6\ 3s, 3p, 3d$.
Calculated Fe~XVIII eigenenergies are compared with experimental
data from the National Institute for Standards and Technology (NIST).
The accuracy of the eigenfunctions is also ascertained by comparing
the Fe~XVIII oscillator strengths for dipole transitions with available data 
from NIST in Table 2.
Photoionization and recombination
calculations both employ the Fe~XVIII eigenfunctions with the same CI.
We carry out two sets of calculations, (i) a 3-level calculations including
only the $n$ = 2 levels, and (ii) the 60-level eigenfunction
expansion including most of the $n$ = 3 complex. 
The inner $2s$-shell excitations are not considered owing to 
computational constraints and possibly weaker couplings (only the allowed
$2s - 3p$ core excitations are likely to be of importance, 
and not the $2s - 3s$ or $2s - 3d$).

For the 3-level case, since we calculate both photoionization and
photo-recombination cross sections, we include many $LS\pi$
symmetries to obtain 15 total $J\pi$ symmetries. Specifically these are
$J = 0 - 7$ for the even parity and $J = 0 - 6$ for the odd parity, and
the $LS\pi$ symmetries used are $L=0-8$ for both singlets and triplets in
both parities.
For the 60-level case, presently we only calculate photoionization
cross section for the ground level and some selected excited levels. Therefore,
we only include $J=0$ for the even ($0^e$) and $J=1$ for the odd parity ($1^o$)
at this moment. All the
$LS\pi$ symmetries that contribute to these two $LS\pi$'s are included.
Of course, for obtaining photo-recombination cross sections we need
to include an many $J\pi$ symmetries as in the 3-level case.

\section{RESULTS AND DISCUSSION}

 The following sections present a sample of the extensive results
from the two sets of calculations for photoionization and
\eion recombination. The 3-level calculations are compared with
earlier theoretical and experimental works. The present close coupling
calculations for the 3-level and the 60-level cases are labeled 3CC and
60CC respectively.

\subsection{Photoionization}

 Fig.~1a presents the BPRM photoionization cross section for the ground level 
$2s^22p^6 \ (^1S_0)$ of Fe~XVII from the 60CC calculation (solid
line), showing the series of resonances converging on to the $n$ = 2
thresholds $2s^22p^5 (^2P^o_{1/2})$, $2s2p^6 (^2S_{1/2})$, and the
$n$ = 3 thresholds. For comparison, the
non-resonant cross sections from a relativistic distorted wave (RDW)
calculation (e.g. \cite{z98}) are also shown (dashed line). The resonance
pattern, and the background cross sections, are essentially similar to 
the 3-level calculations in \cite{metal,hetal} in the relativistic 
random phase approximation (RRPA), with
resonances included using multi-channel quantum defect theory (MQDT),
and the $LS$ and Breit-Pauli $JK$-coupled  R-matrix calculations also
reported by Haque \etal \cite{hetal}.  While there are no significant
differences in magnitude or detail with the earlier calculations, it
might be noted that the near-threshold region spanning the 3 levels of
the $n$ = 2 complex, is rather simple in terms of structure and coupling
effects. Fig.~1b shows an expanded view of this region with series of
resonances $2s^22p^5 \ (^2P^o_{1/2}) \ n \ell$ and the stronger series,
connected to the ground level via a dipole core transition, $2s2p^6 \
(^2S_{1/2}) \ n \ell$. Fig.~2 shows a comparison between the 60CC and
the 3CC calculations, with similar resolution, indicating that below the
$n$ = 2 thresholds there is no significant difference between the two. 

\subsubsection{Channel coupling effects}

 The situation is considerably more complicated above the $n$ = 2 complex.
Although the ground level photoionization cross section of Fe~XVII is not hugely
affected by the $n$ = 3 complexes of resonances (Fig.~1a), the excited
level cross sections are, as seen in Fig.~3a,b,c. This is of considerable
importance in \eion recombination work where photorecombination to group
(A) levels is considered explicitly. The dense and detailed resonance
structures converging on to the 57 $n$ = 3 levels, and in between, would
enhance the effective photoionization and recombination cross sections
and rates far above the background. It might be noted that the resonances
in cross sections below the $n$ = 2, in the energy range covered by the 3-level
calculations shown in Figs.~1 and 2, are much smaller than in the 60CC 
cross sections. This implies, in particular, that resonant recombination into 
the $n$ = 3 series of resonances converging on to a large number of 
excited states will be important for 
\eion recombination, discussed in the next section.

 As the above results show, the 3-level calculations in the present and
earlier works (e.g. \cite{hetal,metal}) are inadequate for the entire energy
range of interest in practical applications for photoionization and
recombination. Also, a 3-level calculation gives little indication of
the complexity of the cross sections, particularly for the excited
states, since it covers only $\Delta n = 0$ core excitations and
couplings that are responsible for resonances. The $\Delta n > 0$
couplings can be much stronger and give rise to more extensive
resonances as in Fig.~3. It is clear that although Fe~XVII is a highly 
charged ion the electron correlation effects are not weak in excited
state photoionization, or in the near-threshold region. Finally, Fe~XVII
is a closed shell system where
simpler approximations (e.g. Haque \etal \cite{hetal}) can be readily 
applied without explicit
consideration of detailed multiplet and fine structure that is more
involved in open-shell systems. Thus 
photoionization of other highly charged ions may not be
amenable to the approximations
described in \cite{hetal}. In fact the atomic structure of
open-shell Fe ions isoelectronic with the third row of the periodic
table present considerable difficulties owing to strong coupling effects
among up to several hundred fine structure levels. Such is the case in a
number of scattering calculations carried out under the Iron Project 
for open-shell Fe ions where extensive resonance structures
manifest themselves \cite{ipwww}.

\subsubsection{Photoexcitation-of-core (PEC) resonances}

 Giant resonances manifest themselves at photon frequencies associated 
with strong dipole transitions in the core ion. These are a particularly
important example of the coupling effects and are called the 
photoexcitation-of-core (PEC) resonances (e.g. \cite{ys87,np91}).
The PEC resonances have the following properties,: 
(i) they are at the photon frequency of the dipole transition in the core, 
(ii) they are present in
photoionization cross sections of the entire Rydberg series  of bound
levels of the \eion system, (iii) their width and height are orders of
magnitude larger than individual Rybderg resonances, and (iv) they are
related to the inverse resonant recombination process DR (discussed in
the next section).
The PEC features are most evident in photoionization
cross sections, as a function of photon energy, of several members of a 
Rydberg series of bound levels
where, in analogy with the DR process, the outer electron is weakly
bound and may be considered a `spectator' interacting but weakly with
the core excitation(s). Fig.~4 shows the large PEC
feature in the 60CC photoionization cross sections of the series $2p^5
\ np \ ^3P_0, n = 3 - 10$ levels of Fe~XVII. The PEC resonances in Fig.~4 
are associated with not just one dipole transition, 
but several transitions belonging to the transition arrays 
$2p^5 - 2p^4 \ 3s,3d$ at about 63 Ry corresponding to all such 
levels included in the 60CC
expansion of the core ion Fe~XVIII (Eq.~1.1, Table 1). The 
PEC resonances rise order of magnitude above the
background, and are much wider than all other resonances. The different
threshold ionization energies in Fig.~4 approach $0$ as $n$ increases.

\subsubsection{Radiation damping of autoionizing resonances}

 Radiation damping of resonances has been addressed in many previous
works. It was pointed out in Ref. \cite{pz97} that it is likely to be of
practical importance only for H-like and He-like core ions, when the
core radiative transition rates are of the same order as the
autoionization rates, typically $10^{13-14} sec^{-1}$, but not for other
ionic systems. Fe~XVII was explicitly mentioned in \cite{pz97} as 
the next possible candidate (other than H-like and He-like ions) for an
investigation of the radiation damping effects {\it in toto}. 
We discuss here the radiation damping involving the resonances associated with
the $n=2$ levels in the context of low-energy recombination, and leave the 
discussion on the $n=3$ resonances in future when the much more extensive 
photo-recombination calculations with the 60CC target  are presented. It
might be noted however that although the radiative decay rates of the
$n$ = 3 levels are higher, they also have additional autoionization
modes of decay into excited $n$ = 2 levels.

The associated core transition rates in Fe~XVIII for $2s^12p^6 \ (^2S_{1/2})$
to $2s^22p^5\ (^2P^o_{3/2})$ and $2s^22p^5 \ (^2P^o_{1/2})$
are $9.13\times 10^{10}$ and $3.31\times 10^{10}$
respectively. Fig.~5 shows an enhanced view of the radiatively damped
(RD) and undamped (NRD) resonances in photoionization of Fe~XVII up to
resonances complexes with $n$ = 16. The PR calculations in the unified
formulation generally employ cross sections up to $n$ = 10 only. 
No significant effect is
discernible between Fig.~5a and 5b, 
and it is concluded that radiation damping of resonances in
$n$-complexes up to $n$ = 10 (e.g. group (A)) is
not likely to affect any practical applications of the computed
photoionization and recombination cross sections. That is not to say
that resonances with sufficiently high $n$ and $\ell$ will not be damped
significantly (or completely); since the autoionization rates decrease
as $n^{-3}$, they must. However, such resonances are extremely narrow
(not, for example, evident in Fig.~5), and do not affect effective
photoionization or recombination cross sections. Therefore it is
unlikely that radiation damping of group (A) resonances (an integral
part of the unified \eion calculations) in any other ionic system up to
the iron-peak elements will be 
important, since the dipole transition probabilities of resonance
transitions in all ions up to Fe
ions are less than or equal to those in Fe~XVIII, of the order of
10$^{12}$ sec$^{-1}$, 
with the already noted exception of H-like and He-like ions \cite{pz97}.

\subsection{Electron-ion recombination}

 Salient features of the \efe1817 recombination are described within the
unified formulation, and with reference to experimental data
from the ion storage ring at Heidelberg, Germany \cite{setal}.

\subsubsection{Comparison with experiment}

 Both the unified cross sections and experimental measurements naturally
measure the combined non-resonant and resonant (RR and DR) contributions
to \eion recombination and should in principle be compared directly. Fig.~6
from Ref. \cite{petal} shows a comparison of the present unified 
cross sections as
computed in detail, and averaged over a gaussian function for comparison
with experiment, together with the experimental cross sections \cite{setal}. 
The agreement is generally very good over the entire range, and in
both the detail and the magnitude of resonances. This includes the
dominant DR contribution below the $^2S_{1/2}$ threshold
connected via a dipole transition; in this range the background
non-resonant contribution (RR) is small.  On the other hand the weaker 
series of resonances, $^2P_{1/2} \ n \ell$ lie in the near-threshold region 
dominated by the non-resonant contribution that rises steeply as $E
\rightarrow 0$, and is therefore not a major contributor to \eion
recombination rate (discussed below).
 Thus the unified theoretical (and experimental) results shown in Fig.~6 
display the three related, but discernible, types of contributions 
to total \efe1817 recombination cross section: 
the overlapping non-resonant (RR) contribution and the DR
contribution from the $^2P^o_{1/2} \ n \ell$ series, and the mainly DR
contribution from the $^2S_{1/2} \ n \ell$ series.

\subsubsection{Resonance strengths and rate coefficients}

 The experimental data in Fig.~6 do not precisely delineate or
identify the resonances, and the background contribution (RR-type) is
not ascertained from the measurements, possibly owing to contribution
from charge transfer \cite{setal}. 
The blended resonance features are fitted to a
beam response function to eliminate the background, and their
energies are determined approximately according to $n$ and $\ell$, using 
non-relativistic $\ell$-dependent quantum defects in the Rydberg formula. 
However, $\ell$ is not a good quantum number and the number of 
$\ell$-resonances within an $n$-complex is not exactly 
known. The theoretical resonances on the other hand are
uniquely identified with the intermediate coupling spectroscopic
designation $(S_iL_iJ_i)n \ell J \pi$. Therefore a 1--1 
correspondence between the experimental measurements and relativistic
cross sections can not be established. Further, since the background
contribution, although dominant at low energies as $E \rightarrow 0$
(Fig.~6), is not considered, a direct comparison with the unified 
cross sections and the experimental data \cite{setal} is not possible. 

 Nonetheless, for $n$-complexes where the
background contribution is small compared to the resonant part, we may
compare the average `resonance strengths' \cite{setal}, although these are not
exactly defined (see \cite{p00} for a definition of the resonance oscillator
strength in terms of the integrated $\frac{df}{d\epsilon}$, the
differential oscillator strength per unit energy).  Figs.~7b,c,d show a
detailed view at high resolution of the first three $^2S_{1/2}n\ell$ complexes,
with $n$ = 6,7,8. In order to ensure complete resolution of resonances
an energy mesh of up to 10$^{-7}$ eV was used before numerical
integration. The integrated, and summed, resonance strengths for the $n$
= 6,7,8 complexes are 1201.2, 421.8, and 221.1 cm$^2$ eV, compared to
experimental values of 1240.2, 412.0, and 253.9 \cite{setal}. The
present value for $n$ = 7 complex is higher than reported in Ref.
\cite{petal} as it is recalculated with higher resolution. The value
for the $n$ = 8 complex has been complemented by the contributions from
$J >$ 7 symmetries, $J\pi$ = 8$^o$ and 9$^e$; without which the value is
200.5 cm$^2$ eV. Although the theoretical resonance strengths were
checked to have converged with
respect to the energy mesh, they seem to be somewhat systematically lower
than the reported experimental data (the theoretical MCBP and the MCDF values
in \cite{setal} also showed the same trend).

Fig.~7a is rather different, in that it shows the much
narrower $^2P^o_{1/2}n\ell$ resonances with high $n$ = 18 --
22. While it is not apparent from the figure, nor from the experimental
data in Fig.~6 (bottom most panel), there are 2,985 resonances found in
the small 0-5 eV range just above ionization threshold of Fe~XVII. The
set of ($E_o, \Gamma_a, \Gamma_r)$ for all these resonances have been 
computed.  The integrated resonance strengths for the $n$ = 18,19 and 20
complexes are 2290.9, 521.6, 290.1 cm$^2$ eV, compared to
experimental values \cite{setal} of 2452.8, 605.7, and 336.5
respectively. Again, we find the integrated values to be up to about
10\% lower than measured, although here the uncertainties are much greater
than for the lower-$n$ $^2S_{1/2} \ n\ell$ resonances in Figs.~7b,c,d
since they may not have been completely resolved and because the
integration energy ranges are very closely spaced. Also, the precise
range of angular momenta $ (\ell, J)$ in the ion storage rings is not
known \cite{setal}, and $J \le$ 7 may not be quite
sufficient for perfect agreement. Detailed comparison of 
resonances beyond this level is neither feasible nor
necessary. Although not all resonances are experimentally identified,
the theoretical cross sections, resonance strengths, and rates (see
below) can be
compared, as shown, to within experimental and theoretical uncertainties.

 Finally we evaluate quantities of practical interest, 
the maxwellian averaged unified
\eion recombination rate coefficients $\alpha_R(T)$ shown in Fig.~8.
These are compared with the sum of the available experimental DR rate
coefficients \cite{setal}, and the background RR-type contribution
extracted from the present theoretical cross sections. The agreement is
of the order of 20\%, the estimated uncertainty in both the experimental
and theoretical datasets. However, we believe that the agreement could
be slightly better since the theoretical results might be somewhat
enhanced if the few $J\pi$ symmetries with $J > 7$, for resonances
with $n \le 10$, are also included. These were omitted to reduce the
complexity of calculations (except to gauge their effect on the
completeness of the $n$ = 8 complex mentioned above), and since their 
resonant contribution is small.

\subsubsection{$\Delta n = 1$ resonances and high energy recombination}

 Thus far only the low energy $\Delta n = 0$
resonances due to the $n = 2$ levels of Fe~XVIII have been considered.
However, as demonstrated in this work, the high energy recombination 
cross sections due to the $n$ = 3 levels are much larger and will be more
important at high temperatures close to the temperature of maximum abundance of
Fe~XVII in collisional equilibrium, around $T = 4-5 \times 10^6$ K. 
This would especially be the case owing to the huge PEC resonances shown
in Fig.~4 that in fact correspond to the peak values of DR due to
resonances converging on to the series limits of strong dipole $\Delta n
= 1$ transitions. Therefore, in addition to the
DR bump corresponding to the $n$ = 2 resonances in the total
$\alpha_R(T)$ shown in Fig.~8, we expect a much larger bump at higher
temperatures from the $n$ = 3 resonances. However, the 60CC calculations
are orders of magnitude more expensive in terms of computational and
other resources and, although they are in progress, would require a 
considerable amount of time to be completed. 

\subsection{Bound states and transition probabilities of Fe~XVII}

 In addition to photoionization and recombination the BPRM calculations
also enable unprecedented quantities of accurate bound state
and transition probability datasets  in intermediate coupling (e.g.
\cite{np99}). The accuracy of these results is comparable to the most
elaborate configuration-interaction atomic structure calculations since
the wavefunction in Eq.~2.1 entails a large configuration
expansion, with each closed channel as a bound configuration of the \eion
system for $E < 0$ and given symmetry $J\pi$. These calculations are in 
progress for all fine structure
levels of Fe~XVII up to $n$ = 10, and all associated E1 $A$- and $f$-values.
Table 3 provides a brief sample of the bound level energies computed for
some of the levels of interest in this work.
Together with the electron impact excitation collision strengths for Fe~XVII in
progress \cite{cp01}, these results should help complete the radiative and 
collisional data for Fe XVII needed for most plasma modeling applications. 

\section{CONCLUSION}
 The most extensive relativistic 
close coupling calculations for photoionization and recombination of an
atomic species are reported for the astrophysically important ion Fe~XVII.
Based on this work we may note the following conclusions: (i)
self-consistent
datasets may be obtained for photoionization and recombination within
the close coupling formulation, (ii)
the coupling to the $n$ = 3 thresholds strongly manifests itself in
excited state photoionization; resonances enhance the effective
cross sections by orders of magnitude particularly below thresholds
coupled via dipole photoexcitation-of-core levels,
(iii) unified \eion
recombination cross sections are in good agreement with
experimental data in terms of both detailed resonance strengths and
rates; the resonances have been delineated at very high resolution 
with considerably more structure than experimentally observed (in
principle all quantum mechanically allowed resonances in intermediate
coupling may be obtained),
(iv) it is necessary to consider the
higher $n$ = 3 levels in high energy recombination that would dominate
the DR part of the \efe1817 recombination at temperatures close to
maximum abundance of Fe~XVII in coronal (collisional) equilibrium, and
(v) it is necessary to account for not only relativistic fine structure
but also the strong coupling among those levels in order to
accurately reproduce the results
for photoionization and recombination over the entire range of practical
interest.

\section{ACKNOWLEDGEMENTS}

 We would like to thank Dr. Werner Eissner for several contributions,
and Mr. Guoxin Chen for assistance with some calculations.
 This work was supported in part by the NSF and NASA. The computational
work was carried out at the Ohio Supercomputer Center in Columbus,
Ohio.

\def\amp{{Adv. At. Molec. Phys.}\ }
\def\apj{{ Astrophys. J.}\ }
\def\apjs{{Astrophys. J. Suppl. Ser.}\ }
\def\apjl{{Astrophys. J. (Letters)}\ }
\def\aj{{Astron. J.}\ }
\def\aa{{Astron. Astrophys.}\ }
\def\aasup{{Astron. Astrophys. Suppl.}\ }
\def\adndt{{At. Data Nucl. Data Tables}\ }
\def\cpc{{Comput. Phys. Commun.}\ }
\def\jqsrt{{J. Quant. Spectrosc. Radiat. Transf.}\ }
\def\jpb{{J. Phys. B}\ }
\def\pasp{{Pub. Astron. Soc. Pacific}\ }
\def\mn{{Mon. Not. R. Astron. Soc.}\ }
\def\pra{{Phys. Rev. A}\ }
\def\prl{{Phys. Rev. Lett.}\ }
\def\zpds{{Z. Phys. D Suppl.}\ }
\def\adndt{At. Data Nucl. Data Tables}

\newpage
\begin{figure}
\centering
\psfig{figure=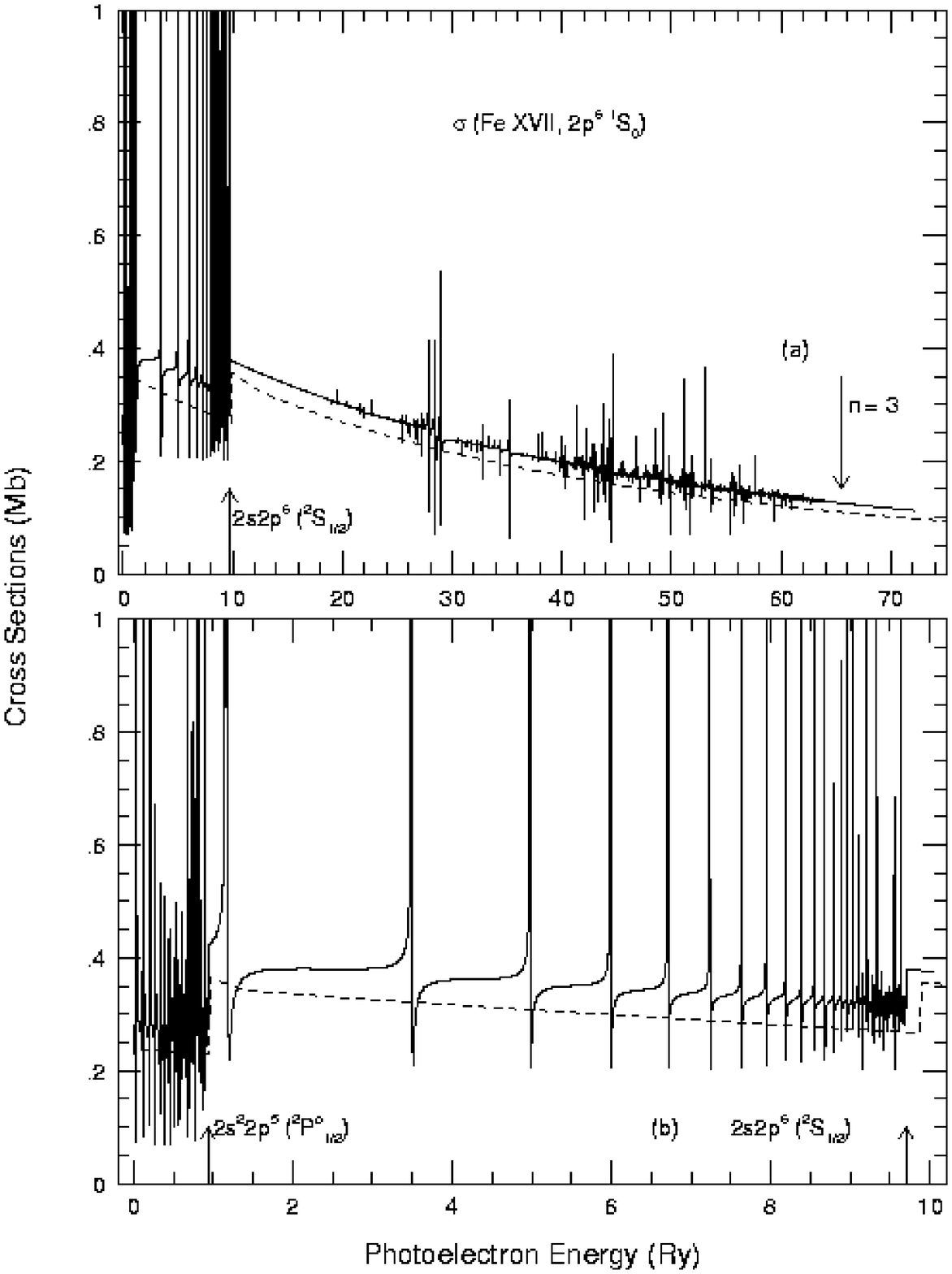,height=17.0cm,width=18.0cm}
\caption{Photoionization cross section of the Fe XVII ground state
$2s^22p^6 \ (^1S_0)$: (a) up to and above the $n$ = 3 thresholds of
the core ion Fe~XVIII, 60-level close coupling (60CC); 
(b) expanded view  up to the $n$ =
2 thresholds $2s^22p^5 \ (^2P^o_{3/2,1/2})$ and $2s2p^6 (^2S_{1/2})$. 
The dashed lines are results from the relativistic distorted
wave calculations \ \protect\cite{z98}.}
\end{figure}

\newpage
\begin{figure}
\centering
\psfig{figure=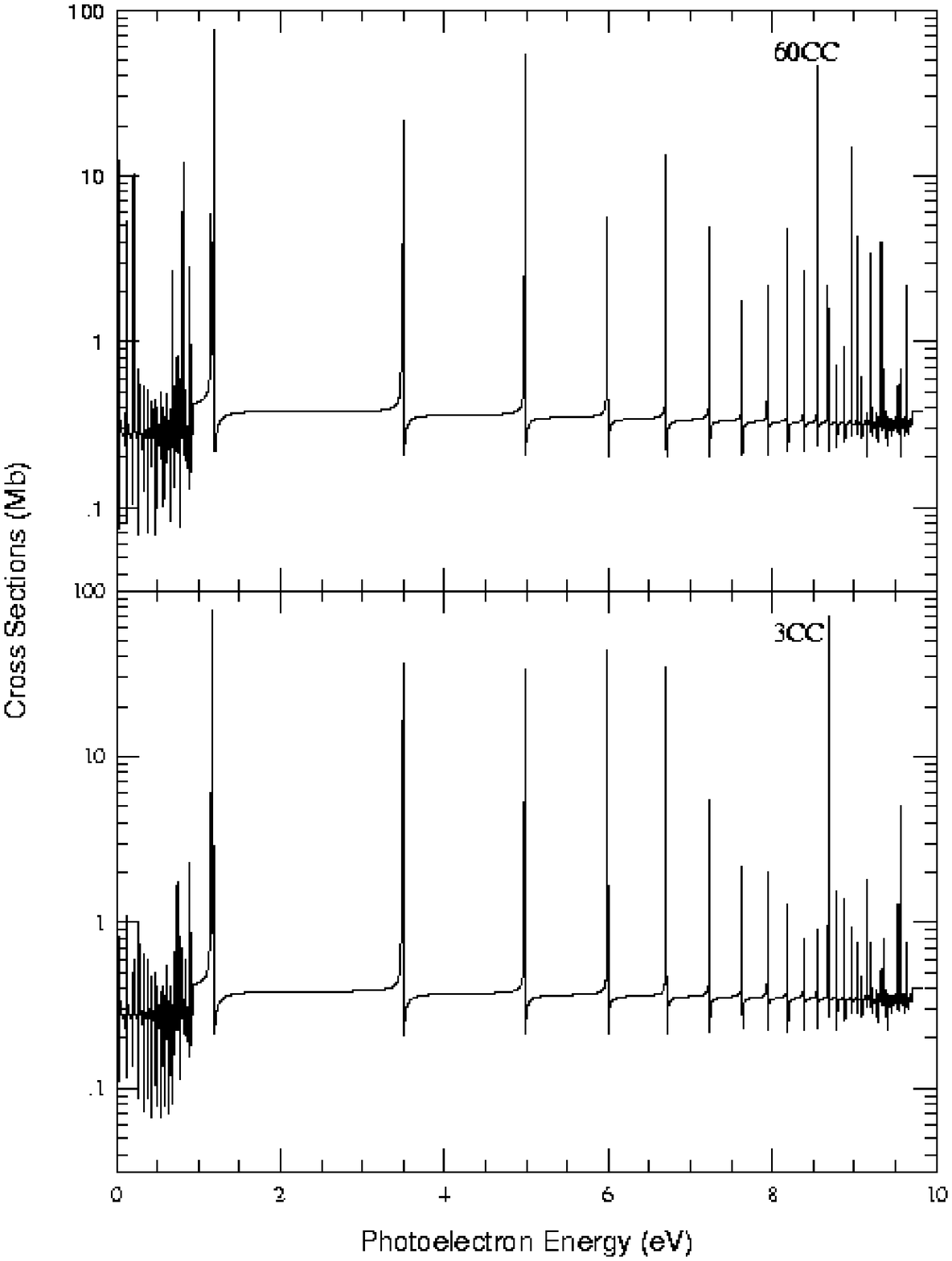,height=17.0cm,width=18.0cm}
\caption{The 60CC and the 3CC cross sections in the
region below the Fe~XVIII $2s2p^6 \ ^2S_{1/2}$ threshold. Although there is no
significant coupling effect below the $n$ = 2 threshold, the $n$ = 3
thresholds are strongly coupled above the $n$ = 2 levels (e.g. Fig.~3).}
\end{figure}

\newpage
\begin{figure}
\centering
\psfig{figure=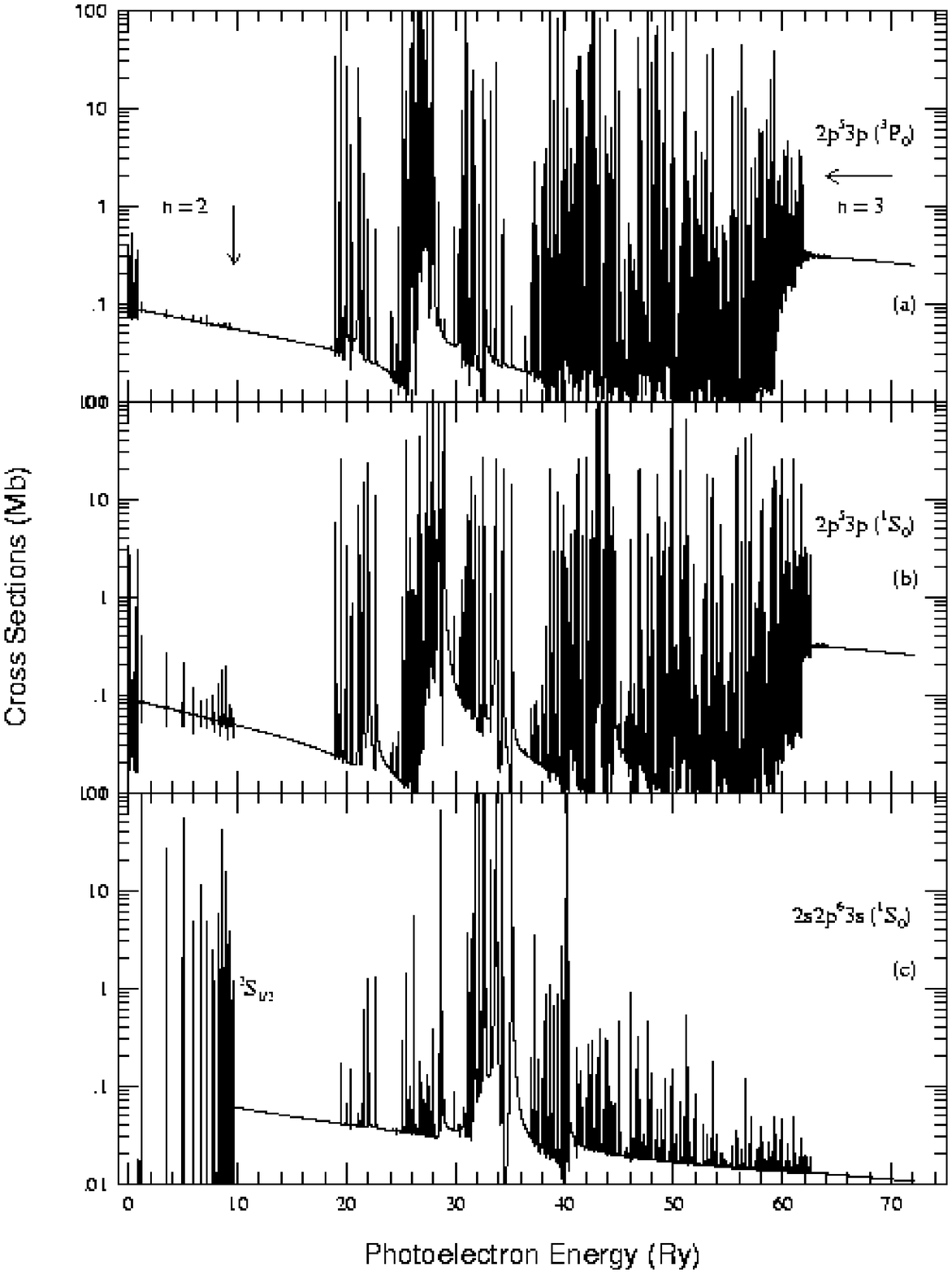,height=17.0cm,width=18.0cm}
\caption{The 60CC photoionization cross sections of the first three excited
levels of the J = 0, even parity symmetry, with extensive series of
resonances converging on to the $n$ = 3 thresholds. The $n$ = 2 resonance
strengths are much weaker.}
\end{figure}

\newpage
\begin{figure}
\centering
\psfig{figure=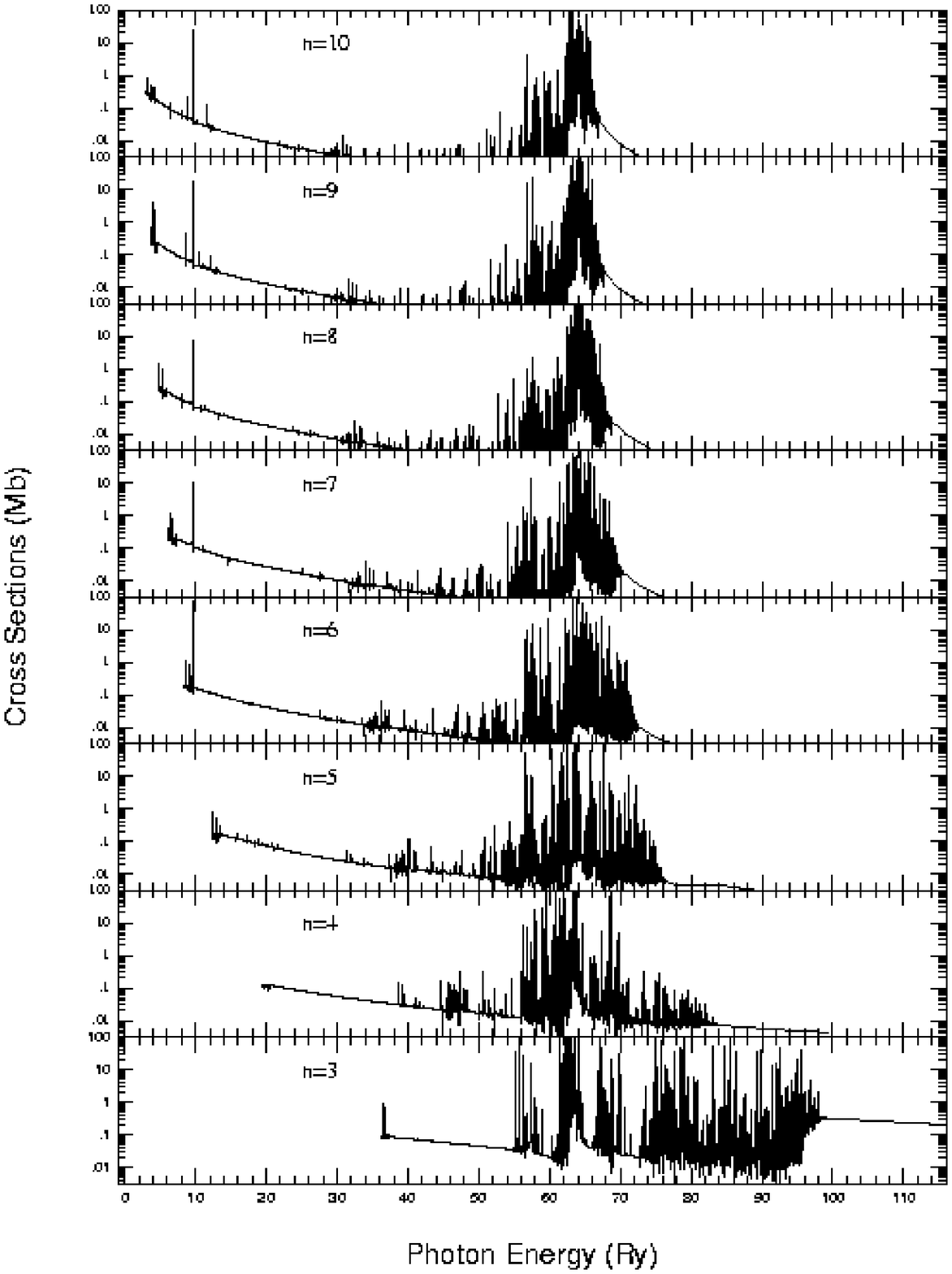,height=17.0cm,width=18.0cm}
\caption{Photoexcitation-of-core (PEC) resonances in photoionization of
the $2p^5 \ np \ (^3P_0)$ Rydberg series of levels of Fe~XVII. The giant
PEC resonance feature at approximately 63 Ry corresponds to strong
dipole excitations in the transition arrays $2p^5 - 2p^4 \ 3s,3d$ within the
Fe~XVIII core (60CC results).}
\end{figure}

\newpage
\begin{figure}
\centering
\psfig{figure=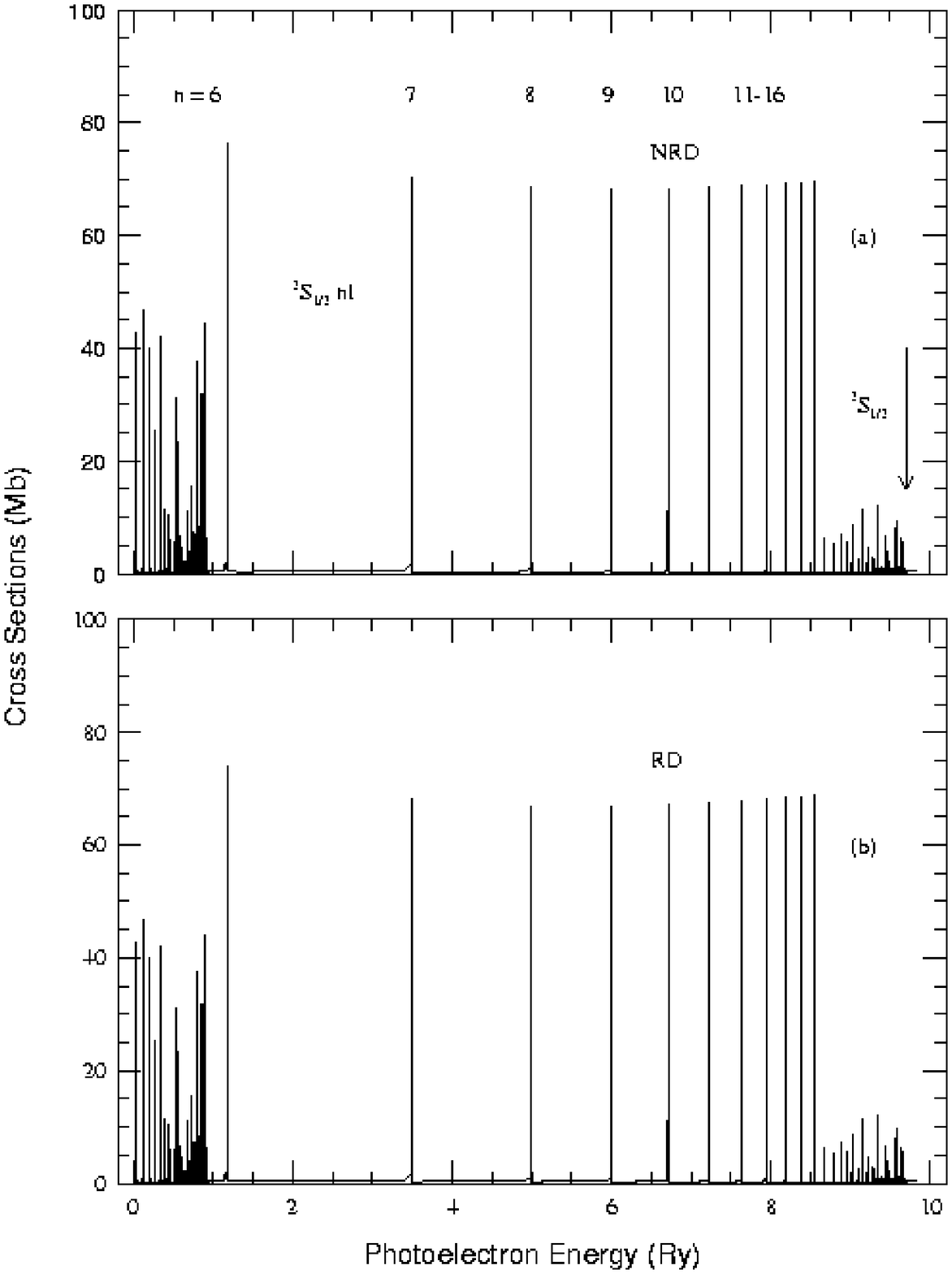,height=17.0cm,width=18.0cm}
\caption{Ground level photoionization cross sections (a) without
radiation damping (NRD) of resonances, and (b) with radiation damping
(RD), showing negligible effect up to the $n$ = 16 complex (higher
$n$-complexes are not resolved).}
\end{figure}

\newpage
\begin{figure}
\centering
\psfig{figure=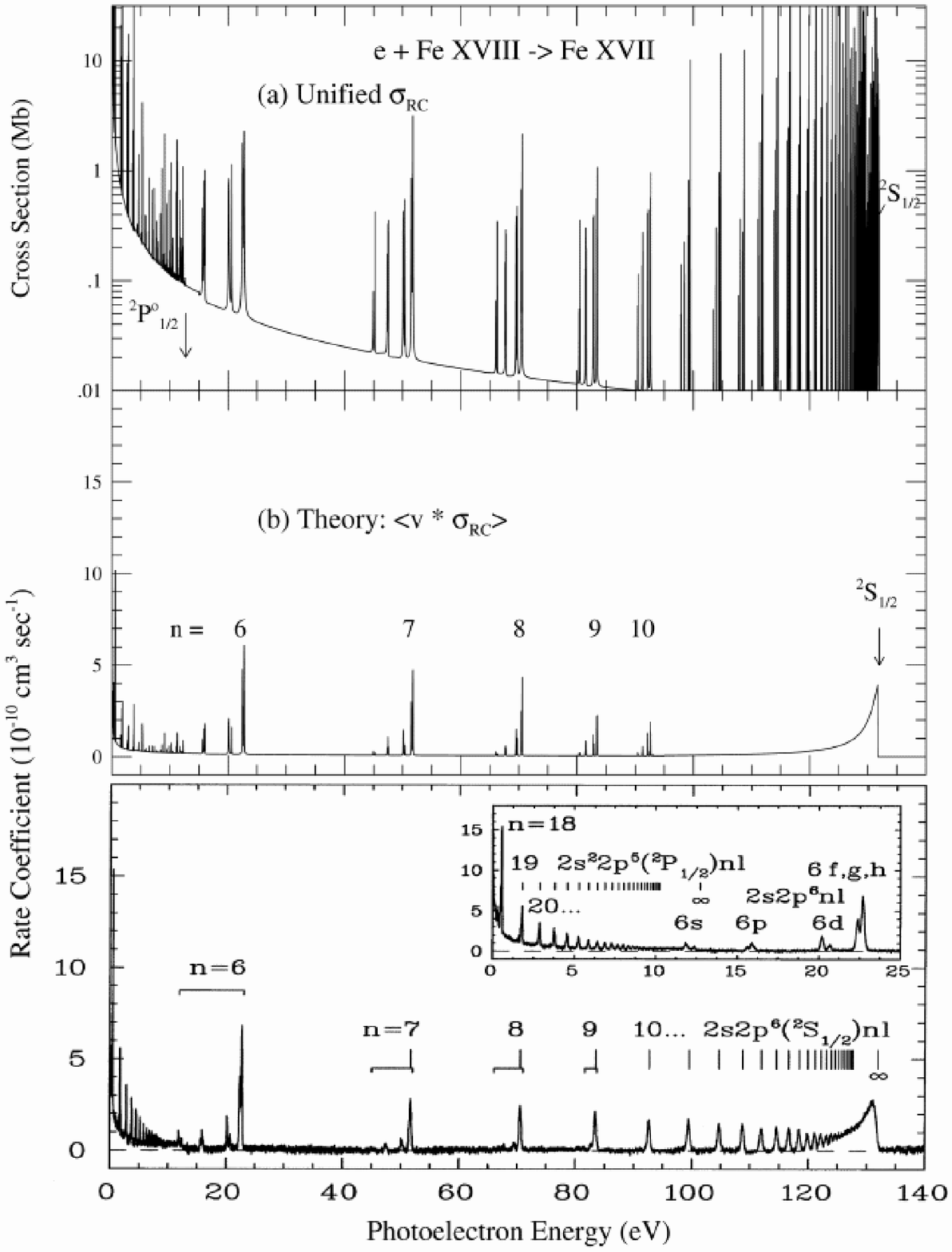,height=17.0cm,width=18.0cm}
\caption{Unified (e + Fe~XVIII) $\rightarrow$ Fe~XVII recombination
cross sections (upper panel) with detailed resonance complexes below the
n = 2 thresholds of Fe~XVIII; gaussian
averaged over a 20 meV FWHM (middle panel); experimental data from ion
storage ring measurements \ \protect\cite{setal}(bottom panel).}
\end{figure}

\newpage
\begin{figure}
\centering
\psfig{figure=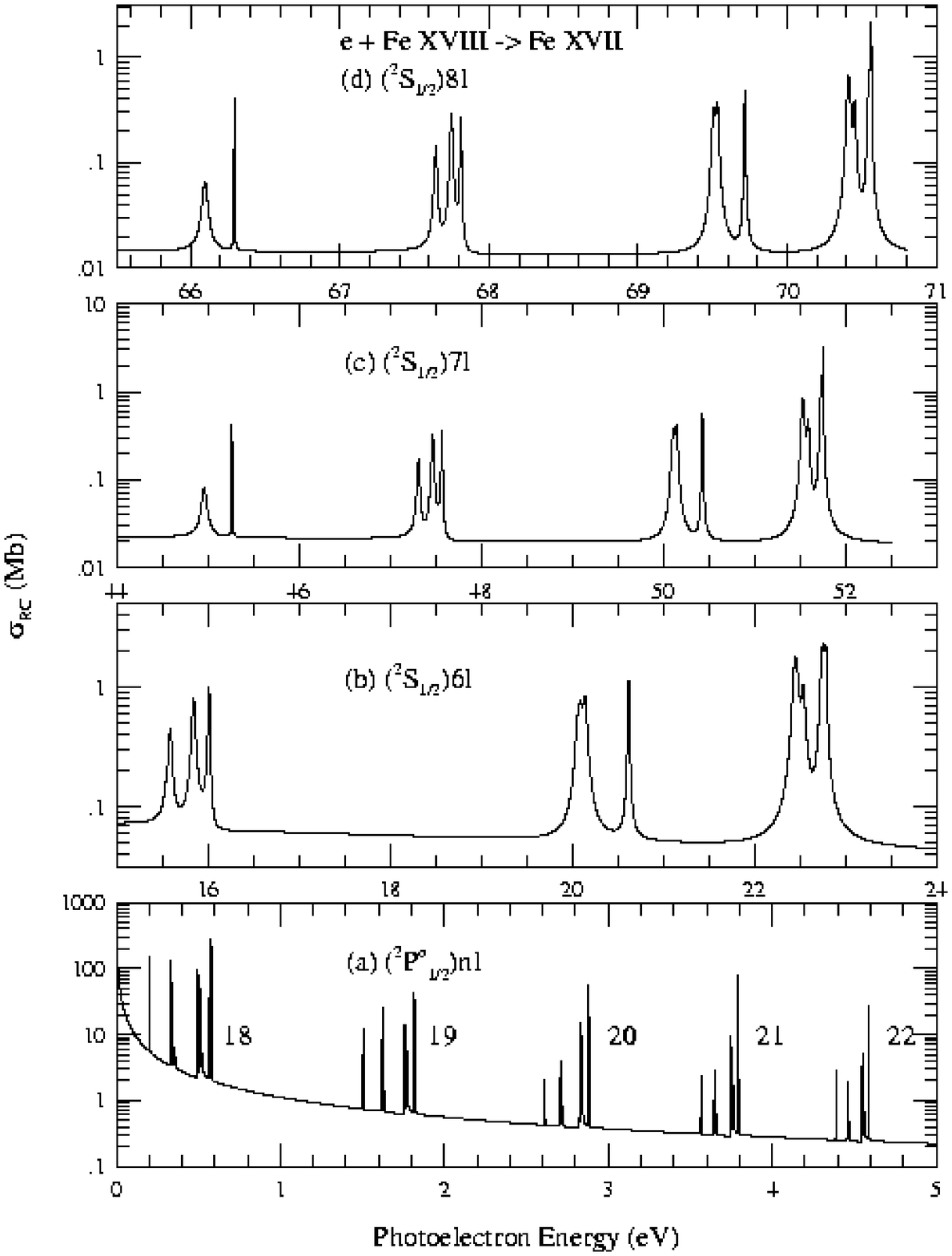,height=17.0cm,width=18.0cm}
\caption{Resolved resonance complexes in the unified 
recombination cross sections: (a) $^2P^o_{1/2} \ n\ell$, 
(b) $^2S_{1/2} \ 6\ell$, (c) $^2S_{1/2} \ 7\ell$, (8)$^2S_{1/2} \
8\ell$. The lowest resonance group in the $n$ = 6 complex lies below the
$^2P^o_{1/2}$ threshold among high-$n$ $^2P^o_{1/2} \ n\ell$ resonances and
is not shown. The complexity of resonance structures is barely apparent 
from figure (a), and even less so from the experimental data in Fig.~6 
(lowest panel); 2,985 resonances have been resolved in the 0-5 eV range
shown in (a).}
\end{figure}

\newpage
\begin{figure}
\centering
\psfig{figure=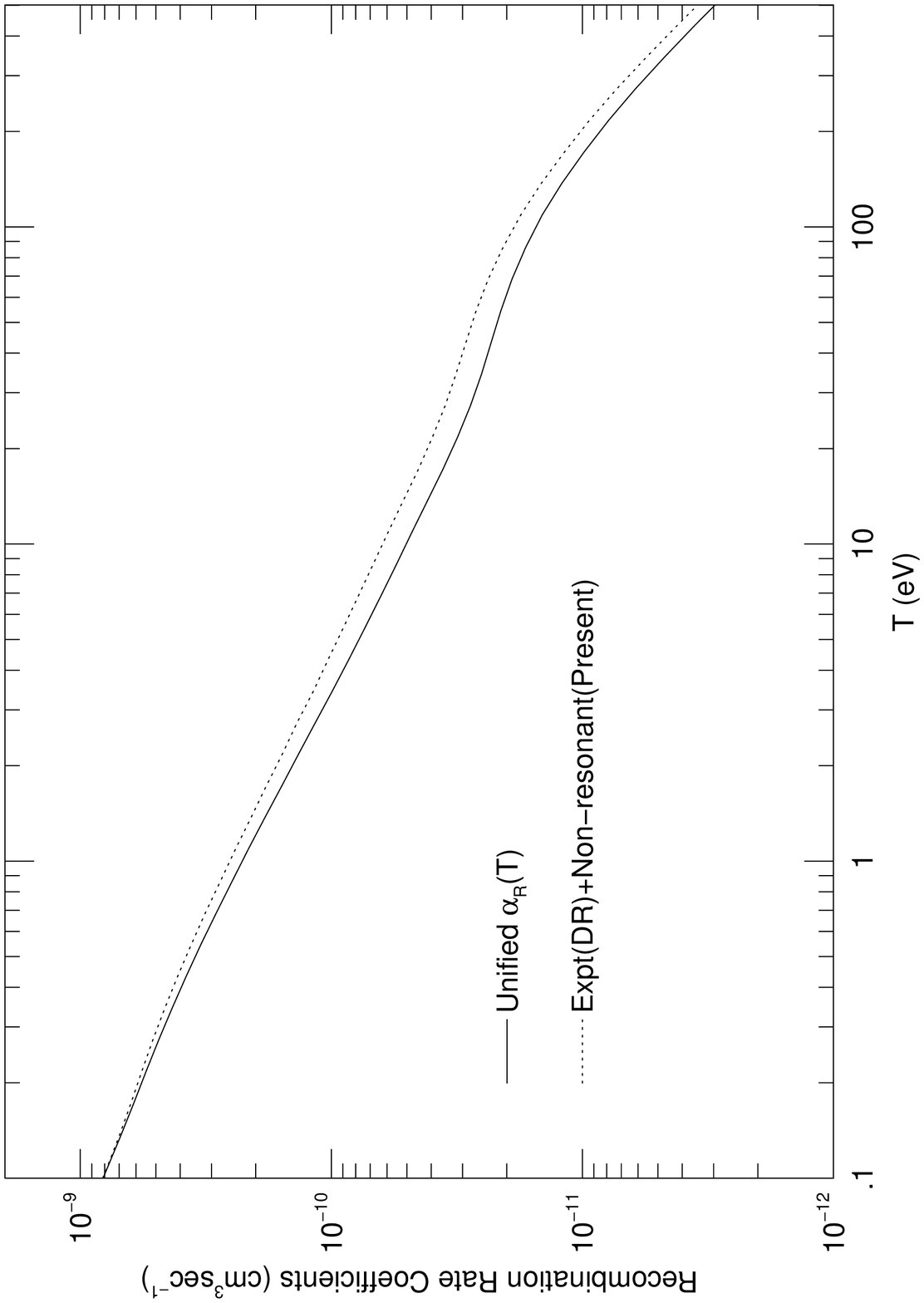,height=17.0cm,width=18.0cm}
\caption{Comparison of the low-temperature unified recombination rate 
coefficient
$\alpha_R(T)$ with $\sigma_{RC}$ including $n$ = 2 resonances only (as in
the experimental data \ \protect\cite{setal}),
and the sum of the experimental DR rate coefficient +
non-resonant background (RR-type) contribution extracted from the present
calculations. The maximum difference is about 20\%.}
\end{figure}

\newpage
\begin{table}
\caption{Fine structure energy levels for the 60CC eigenfunction expansion
of the target ion Fe~XVIII. The level energies are in Ry.
\label{table1}}
\begin{tabular}{rccccccrccccc}
 i & Configuration  &Term  &2$J$&  $E$(Present) & $E$(NIST) &\hspace*{0.3in} &
 i & Configuration  &Term  &2$J$&  $E$(Present) & $E$(NIST) \\
\noalign{\smallskip}
\tableline
\noalign{\smallskip}
 1 &   $2s^22p^5$   &$^2P$ & 3  &    0.00000 &  0.      & &
31 &   $2s^22p^43d$ &$^4D$ & 5  &   62.299         \\
 2 &                &$^2P$ & 1  &    0.94212 &  0.93477      & &
32 &                &$^4D$ & 7  &   62.311         \\
 3 &   $2s  2p^6$   &$^2S$ & 1  &    9.80691 &  9.70228      & &
33 &                &$^4D$ & 1  &   62.429   & 62.906      \\
 4 &   $2s^22p^43s$ &$^4P$ & 5  &   56.991   & 56.690      & &
34 &                &$^4D$ & 3  &   62.341   & 63.050      \\
 5 &                &$^2P$ & 3  &   57.239   & 56.936      & &
35 &   $2s^22p^43p$ &$^2P$ & 3  &   62.461         \\
 6 &                &$^4P$ & 1  &   57.671   & 57.502      & &
36 &   $2s^22p^43d$ &$^4F$ & 9  &   62.535         \\
 7 &                &$^4P$ & 3  &   57.836   & 57.572      & &
37 &                &$^2F$ & 7  &   62.629         \\
 8 &                &$^2P$ & 1  &   58.068   & 57.798      & &
38 &   $2s^22p^43p$ &$^2P$ & 1  &   62.686         \\
 9 &                &$^2D$ & 5  &   58.609   & 58.000      & &
39 &   $2s^22p^43d$ &$^4P$ & 1  &   62.767   & 62.496      \\
10 &                &$^2D$ & 3  &   58.642   & 58.355      & &
40 &                &$^4P$ & 3  &   62.905   & 62.625      \\
11 &   $2s^22p^43p$ &$^4P$ & 3  &   59.209   &             & &
41 &                &$^4F$ & 5  &   62.985         \\
12 &                &$^4P$ & 5  &   59.238   &             & &
42 &                &$^2P$ & 1  &   63.123         \\
13 &                &$^4P$ & 1  &   59.478   &             & &
43 &                &$^4F$ & 3  &   63.156         \\
14 &                &$^4D$ & 7  &   59.525   &             & &
44 &                &$^2F$ & 5  &   63.177   & 62.698      \\
15 &                &$^2D$ & 5  &   59.542   &             & &
45 &                &$^4F$ & 7  &   63.271         \\
16 &   $2s^22p^43s$ &$^2S$ & 1  &   59.947   & 59.916      & &
46 &                &$^2D$ & 3  &   63.302         \\
17 &   $2s^22p^43p$ &$^2P$ & 1  &   59.982   &             & &
47 &                &$^4P$ & 5  &   63.451   & 62.911      \\
18 &                &$^4D$ & 3  &   60.005   &             & &
48 &                &$^2P$ & 3  &   63.574   & 63.308      \\
19 &                &$^4D$ & 1  &   60.012   &             & &
49 &                &$^2D$ & 5  &   63.672   & 63.390      \\
20 &                &$^2D$ & 3  &   60.147   &             & &
50 &                &$^2G$ & 7  &   63.945         \\
21 &                &$^4D$ & 5  &   60.281   &             & &
51 &                &$^2G$ & 9  &   63.981         \\
22 &                &$^2P$ & 3  &   60.320   &             & &
52 &                &$^2S$ & 1  &   64.198   & 63.919      \\
23 &                &$^2S$ & 1  &   60.465   &             & &
53 &                &$^2F$ & 5  &   64.200         \\
24 &                &$^4S$ & 3  &   60.510   &             & &
54 &                &$^2F$ & 7  &   64.301         \\
25 &                &$^2F$ & 5  &   60.851   &             & &
55 &                &$^2P$ & 3  &   64.432   & 64.138      \\
26 &                &$^2F$ & 7  &   61.028   &             & &
56 &                &$^2D$ & 5  &   64.488   & 64.160      \\
27 &                &$^2D$ & 3  &   61.165   &             & &
57 &                &$^2D$ & 3  &   64.703   & 64.391      \\
28 &                &$^2D$ & 5  &   61.272   &             & &
58 &                &$^2P$ & 1  &   64.767   & 64.464      \\
29 &                &$^2P$ & 3  &   61.761   &             & &
59 &                &$^2D$ & 5  &   65.481   & 65.305      \\
30 &                &$^2P$ & 1  &   61.899   &             & &
60 &                &$^2D$ & 3  &   65.669   & 65.468      \\
\end{tabular}
\end{table}

\newpage

\begin{table}
\caption{Comparison of the present dipole oscillator strengths 
(the $gf$ values) for fine-structure transitions in Fe~XVIII with
the NIST compiled data. See Table I for the level index.
\label{table2}}
\begin{tabular}{rrcc}
i & j & NIST & Present \\
\noalign{\smallskip}
\tableline
\noalign{\smallskip}
1 & 3 & 0.242 & 0.222 \\
1 & 4 & 0.021 & 0.018 \\
1 & 6 & 0.015 & 0.012 \\
1 & 8 & 0.104 & 0.079 \\
1 & 9 & 0.242 & 0.191 \\
1 &16 & 0.019 & 0.014 \\
1 &43 & 0.096 & 0.085 \\
1 &52 & 0.975 & 0.879 \\
1 &55 & 2.300 & 2.344 \\
1 &57 & 0.516 & 0.511 \\
1 &59 & 0.193 & 0.190 \\
1 &60 & 0.011 & 0.008 \\
2 & 3 & 0.107 & 0.100 \\
2 & 8 & 0.116 & 0.095 \\
2 &10 & 0.196 & 0.167 \\
2 &16 & 0.079 & 0.055 \\
2 &52 & 0.169 & 0.159 \\
2 &55 & 0.399 & 0.373 \\
2 &57 & 1.855 & 1.579 \\
2 &60 & 1.794 & 1.831 \\
\end{tabular}
\end{table}

\newpage

\begin{table}
\caption{Comparison of the present energies and the NIST values (Ry)
for selected fine structure levels of Fe~XVII.
\label{table3}}
\begin{tabular}{lrr}
 Level & Present & NIST \\
\noalign{\smallskip}
\tableline
\noalign{\smallskip}
$2s^2 2p^6\ ^1S_0 $      &   0.0000  &  0.0000  \\ 
$2s^2 2p^5 3p\ ^3P_0 $   &  56.6532  & 56.5155  \\
$2s^2 2p^5 3p\ ^1S_0 $   &  58.0986  & 57.8897  \\ 
$2s^2 2p^5 3s\ ^1P^o_1 $ &  53.5708  & 53.4300 \\ 
$2s^2 2p^5 3s\ ^3P^o_1 $ &  54.4475  & 54.3139 \\
$2s^2 2p^5 3d\ ^3P^o_1 $ &  59.1231  & 58.9810 \\ 
$2s^2 2p^5 3d\ ^3D^o_1 $ &  59.8763  & 59.7080 \\ 
$2s^2 2p^5 3d\ ^1P^o_1 $ &  60.8849  & 60.6000 \\ 
$2s 2p^6 3p\ ^1P^o_1 $   &  65.7990  & 65.6010 \\
$2s 2p^6 3p\ ^3P^o_1 $   &  66.1216  & 65.9230 \\ 
$2s^2 2p^5 4s\ ^1P^o_1 $ &  72.0294  & 71.8600 \\
$2s^2 2p^5 4s\ ^3P^o_1 $ &  72.9394  & 72.7400 \\ 
$2s^2 2p^5 4d\ ^3P^o_1 $ &  74.1757  & 73.9400 \\ 
$2s^2 2p^5 4d\ ^3D^o_1 $ &  74.4979  & 74.3000 \\ 
$2s^2 2p^5 4d\ ^1P^o_1 $ &  75.3539  & 75.1700 \\ 
\end{tabular}
\end{table}


\end{document}